\documentclass[12pt]{article}
\usepackage[svgnames]{xcolor}

\newcommand{\beq}{\begin{equation}}
\newcommand{\eeq}{\end{equation}}
\newcommand{\beqa}{\begin{eqnarray}}
\newcommand{\eeqa}{\end{eqnarray}}
\newcommand{\beqar}{\begin{eqnarray*}}
\newcommand{\eeqar}{\end{eqnarray*}}

\newcommand{\eps}{\epsilon}

\newcommand{\inn}{\!\cdot\!}

\newcommand{\labell}[1]{\label{#1}} 
\newcommand{\reef}[1]{(\ref{#1})}
\newcommand\prt{\partial}

\newcommand\cF{{\cal F}}

\newcommand\cA{{\cal A}}
\newcommand\cE{{\cal E}}
\newcommand\cM{{\cal M}}

\newcommand\cL{{\cal L}}

\newcommand\cB{{\cal B}}

\newcommand\Tr{{\rm Tr}}

\begin{document}

\vspace*{1cm}

\begin{center}
{\bf \Large  S-dual Amplitude and  $D_3$-Brane Couplings }\\
\vspace*{1cm}

  Komeil Babaei Velni\footnote[1]{babaeivelni@guilan.ac.ir} and  H. Babaei-Aghbolagh\footnote[2]{h.babaei@uma.ac.ir}\\

\vspace*{1cm}
$^{1}${Department of Physics, University of Guilan,\\ P.O. Box 41335-1914, Rasht, Iran}
\\
 \vspace{0.5cm}
$^{2}${Department of Physics, University of Mohaghegh Ardabili, \\ P.O. Box 179, Ardabil, Iran}
\\
\vspace{2cm}



%
%
%
%

\end{center}





\begin{abstract}
\baselineskip=18pt
Recently, it has been observed that the $IIB$ scattering amplitudes are compatible with the standard rules of S-duality.
 Inspired by this observation, we will find the tree-level S-matrix elements of one Ramond-Ramond and three open strings by imposing this symmetry on the tree-level S-matrix elements of one Kalb-Ramond and three open strings.
We also find a $SL(2,R)$ invariant form of the $D_3$-brane effective action containing four gauge fields with derivative corrections that was derived from one-loop level four-point amplitude. Using the expansion of the nonlinear $SL(2,R)$ invariant structures, we find the action with derivative corrections at the level of more gauge fields.  

\end{abstract}
\vskip 0.5 cm


\vfill
\setcounter{page}{0}
\setcounter{footnote}{0}
\newpage

\section{Introduction} \label{intro}
It has been shown that the consistency of an electrodynamics theory under the duality transformation can be expressed as a requirement in which the Lagrangian must transform under duality in a particular way, defined by the Noether-Gaillard-Zumino (NGZ) identity \cite{Gaillard:1981}. So, in general, a nonlinear electrodynamics theory that saturates the NGZ identity, is a self-dual theory. One of the familiar self-dual electrodynamics theory is the Born-Infeld theory $\cL_{BI}=\sqrt{-\det(\eta_{ab}+F_{ab})}-1$, where $F_{ab}=\partial_{a}A_{b}-\partial_{b}A_{a}$ is gauge field strength and $A$ is polarization of the gauge field.

 This theory could be enjoyed the SL(2,R) S-duality (relevant to string theory which insinuates a strong-weak coupling duality of such theories \footnote{From the presence of two two-form gauge fields $B_2$ ($NSNS$ two-form) and 
 $C_2$ ($RR$ two-form) in the string theory, a string can carry two types of charge. These two-forms form a doublet of $SL(2,R)$ it follows that the string
also transform as a doublet. On the other hand, the transformation of the complex field $\tau=C_0+ie^{\phi_0}\rightarrow -1/\tau $ evaluated at $C_0 = 0$, changes the sign of the dilaton, which implies that the string coupling constant maps to its inverse \cite{Becker:2007}.}) after taking axion and dilaton fields into account. This action is not invariant under the S-duality, however, its equations of motion and energy-momentum tensor are invariant under the S-duality as it was shown for the electric-magnetic duality\cite{Green:1996,komeilhosein}. 

The supersymmetric p-branes that supported by Ramond-Ramond $(RR)$ sources are the solutions of type $II$ superstring theories.
These branes have alternative description in terms of open strings with Dirichlet boundary conditions.
$D$-branes  are described by an effective Dirac-Born-Infeld $(DBI)$ action (supplemented with extra couplings to RR fields) which is closely connected to the Born-Infeld ($BI$) type effective action of open string theory. We consider the $D_3$-brane of type $IIB$ theory which play a special role in this theory. The action corresponding  to this brane in the Einstein frame can be interpreted as a generalization of four dimensional $BI$ action coupled to a special background metric, dilaton, axion, etc \cite{Polchinski}.
\beqa
S_{D3}=-T_{D3}\int {d^4x}\sqrt{-det(g_{ab}+e^{-\phi/2}B_{ab})}+T_{D3}\int [C^{(4)}+C^{(2)} B+\frac{1}{2}C^{(0)} B B],
\eeqa
where all the bulk fields in the action are pull-back onto the world-volume of $D$-brane.

 It has been demonstrated that the $D_3$-brane action (combined with type IIB effective action) and the corresponding equation of motion and energy-momentum tensor are invariant under the $SL(2, R)$ symmetry of type IIB theory\footnote{This symmetry is not shared by the full type $IIB$ superstring theory. Indeed, it is broken by
a variety of stringy and quantum effects to the infinite discrete subgroup $SL(2, Z)$.} \cite{Green:1996}. In fact, this action satisfies the NGZ identity.
As was pointed out in \cite{Roiban:2012} that the duality invariance of Hamiltonian  and thus of the corresponding energy momentum tensor should imply the invariance of the S-matrix.



It is known that the S-matrix elements satisfy the Ward identity corresponding to the S-duality\cite{Garousi:2017}.
At the level of the equations of motion, $D_3$-brane effective action is invariant under nonlinear S-duality transformation \cite{Green:1996}. So at higher orders, one dose not expect
the effective action to be consistent with the nonlinear S-duality.

It was pointed in \cite{Garousi:1108} that the $BI$ action involving on-shell gauge field is invariant  under the linear S-duality up to $F^4$ terms. 
From this point, the disk-level scattering amplitude of four gauge fields on the world volume of a single $D_3$-brane, which appears as a contact term, should be invariant under the linear S-duality. The leading order S-matrix element of the six gauge fields has both contact terms coming from the $F^6$ terms in the $BI$ action and massless poles coming from the $F^4$ terms in the $BI$ action. So as expected, no part of this amplitude is not invariant under the linear $SL(2,R)$ transformation separately, however, the combination of these two parts is invariant\cite{hosein:1304}. It has been discussed in \cite{Garousi:1108} that by imposing the linear S-duality transformation on the S-dual invariant amplitude of four gage fields, the pole part of the amplitude of six gage fields could be constructed.  Then, $F^6$-contact-terms could be found by applying the linear S-duality again. 
 Then using the linear S-duality, one may find the contact terms of $F^8$, and so on. This result confirms the  $BI$ action with abelian gage fields as the effective action of a single D-brane.

So, in order to investigate the behavior of S-matrix elements under duality transformation, it is convenient to  separate the $\alpha'$ expansion of them into two parts: contact term and massless $n$-poles  
\beqa
\cA= \cA_{contact}+\sum_n A_{n-poles}.
\eeqa
Neither $\cA_{contact}$ nor $A_{n-poles}$ are invariant under the linear S-duality. In fact, the combination of these two parts, \i.e. $\cA$, must be invariant. On the other hand, the pole part of amplitude may transform to contact part under the linear S-duality transformation, so the effective action, that is made of contact part, satisfies the Ward identity corresponding to S-duality\footnote{On the other hand, the T-duality transformation dose not transform the pole terms to contact terms, so the contact terms always satisfy the Ward identity corresponding to T-duality \cite{komeil}.}.
In the case that the S-matrix elements have no massless poles, one can find the corresponding couplings from the consistency of them with the dualities\cite{Garousi:1511}.

Unlike the gauge field transformation that is carried by its field strength, all other transformations are done by field potential. So, the pole part of S-matrix dose not transform to contact part under the closed string $SL(2,R)$ transformation. It could be noted that if the S-matrix involves the antisymmetric $NSNS$ ($B$-field) closed string, it could be found the amplitude in terms of $B$-field strength $H$ by combining the pole part in amplitude which are produced by the gauge fields and some of the contact terms which are produced by replacement $F\rightarrow F + B$ \cite{Garousi:2017}.

The S-matrix elements of one $RR$ two-form field and three gauge fields that appears as a pole term and the S-matrix elements of one $B$ field and three gauge fields that contains a contact term as well as a pole term have been calculated in \cite{hosein:1304}. It has been shown that the combination of these two amplitudes could be appeared in terms of a $SL(2,R)$ invariant structure. In this paper we show that these amplitudes transform to each other under the linear S-duality transformation
\beqa
\cA_{1-pole}(C^2,F^3)\rightarrow \cA_{contact}(B,F^3)+\cA_{1-pole}(B,F^3).\labell{AB}
\eeqa
Inspired by the above observation that the S-matrix elements of one closed string  and three open strings should satisfy the Ward identity corresponding to the S-duality, it could be proposed the following expression for the massless n-pole amplitude of n $RR$ two-form.
\beqa
\cA_{n-pole}((C^2)^n,F^{4-n})\rightarrow  \cA_{contact}(B^n,F^{4-n})&+&\cA_{1-pole}(B^n,F^{4-n})+\cdots +\cA_{n-pole}(B^n,F^{4-n})\nonumber
\eeqa
where $n=0,1,2,3,4$. The contact part of above amplitude could be exactly reproduced by the replacement $F\rightarrow F + B$ in the $D_3$-brane effective action.

It is convenient to consider the normalization factor $2\pi \alpha'$ in front of the gauge field in $DBI$ action. With this normalization, the $DBI$ action is at the leading
order of $\alpha'$.  The $\alpha'$ corrections to Born-Infeld action have been studied in \cite{Abouelsaood:1986gd,Tseytlin:1987ww,Andreev:1988cb, Wyllard:2000qe, Andreev:2001xx, Wyllard:2001ye} in the $\sigma$-model approach.


The derivatives of gauge fields could be appeared in $D_3$-brane effective action. The study of the behavior of four point function under  the linear S-duality could be extended to four point function with derivative corrections\cite{Carrascoa}. It has been shown in \cite{Chemissany} that the electromagnetic self-duality holds for such functions with derivative corrections at the order of $\alpha'^2$. So, one might expect that the $SL(2,R)$ S-dual invariant structure could be appeared for each order of $\alpha'$ independently.  On other words, we have to answer the following question that should S-duality be a symmetry of full $D_3$-brane action with all derivative corrections? There is no exact reasoning for this, but there is still an alternative (albeit somewhat exploratory) logic suggesting that $D_3$-brane action (but not actions for other D-branes!) should be S-duality covariant\footnote{We would like to thank A.A.Tseytlin for discussions on this point.} \cite{Green:1996} (see also \cite{Tseytlin8}).

By calculating the scattering amplitude of four open string in $IIB$ superstring theory, it has been found that the leading order of $D_3$-brane effective action in superstring theory contains four gauge fields and four derivatives\cite{Wyllard:2000qe,Shmakova,Chemissany}
\beqa
L_{(\partial F{})^{4}}&=& e^{-2\phi}\,\,\bigg(\frac{1}{4} \partial_{c}F^{ef} \partial^{c}F^{ab}\partial_{d}F_{ef} \partial^{d}F_{ab} -  \frac{1}{2} \
\partial_{c}F_{df} \partial^{c}F^{ab} \partial_{e}F_{b}{}^{f}  \partial^{e}F_{a}{}^{d}\nonumber\\
&& -  \partial_{c}F_{a}{}^{d}\partial^{c}F^{ab} \partial_{f}F_{de} \partial^{f}F_{b}{}^{e} + \frac{1}{8} \partial_{c}F_{ab} \partial^{c}F^{ab} \
\partial_{f}F_{de} \partial^{f}F^{de}\bigg).\labell{LF4}
\eeqa
This action  satisfies the NGZ identity, and is consistent with electro-magnetic self-duality\cite{Chemissany}. We will find the form of above action that is manifestly S-dual. In fact, we will find the above action in terms a nonlinear S-dual structure. From the nonlinear expansion of the $SL(2,R)$ invariant structures, we find this effective action up to level of $N$ gauge fields and four derivatives.

The outline of the paper is as follows: We begin in section 2 by studying the S-duality transformations of bosonic fields and find some  $SL(2,R)$ invariant structures involving these fields. In section 3, we show that the compatibility of the S-matrix elements of one gauge field, two scalar fields and one $RR$ field $C^{(2)}$ with the S-duality
generates the S-matrix elements of one gauge field, two scalar fields and one $NSNS$ $B$-field. We find a S-dual invariant form of these amplitudes and then predict the axion amplitude. In section 4, we calculate the S-matrix elements of three gauge fields and one $NSNS$ B-field by applying the S-duality transformation on the S-matrix elements of three gauge fields and one $RR$ field $C^{(2)}$. In section 5, using $SL(2,R)$ invariant structures found in section 2, we write the $D_3$-brane effective action involving four gauge fields and four derivatives in S-dual form. This expression could be extended to include more gauge fields.

\section{Nonlinear SL(2,R) structure}

The axion $C_0$ combines with the dilaton $\phi_0$ to give the complex scalar field $\tau=C_0+ie^{-\phi_0}$. Now consider the matrix $\cM$
\beqa
 {\cal M}=e^{\phi_0}\pmatrix{|\tau|^2&C_0 \cr C_0&1}.\labell{M}
\eeqa
The nonlinear $SL(2,R)$ transformation of the complex scalar field and the above matrix are given by\cite{Gibbons:9509}
\begin{eqnarray}
\tau\rightarrow \frac{p\tau+q}{r\tau+s}\,\,\,\,\,\,\,,\,\,\,\,\,\,{\cal M}\rightarrow \Lambda {\cal M}\Lambda ^T   \,\,\,\,\,\,; \,\,\,\,\,\, \,\,\,\,\,\,\Lambda=\left(
\begin{array}{cc}
p&q  \\
r&s
\end{array}
\right)\in\, {SL(2,R)}.\labell{tM}
\end{eqnarray}
 For the gauge field we have the following transformation
\beqa
F_{ab}&\rightarrow &sF_{ab}+r\tilde{G}_{ab}, \nonumber\\
G_{ab}&\rightarrow &pG_{ab}-q\tilde{F}_{ab},
\eeqa
 
where ${F}_{ab}=-\frac{1}{2} \epsilon_{abcd}\tilde{F}^{cd}$, the antisymmetric tensor $G_{ab}$ is given by $-2\partial \cL(F)/\partial F^{ab}$ and $\cL$ is the Lagrangian. By considering a gauge field doublet $\cF_{a b}$ containing $F_{a b}$ and $G_{a b}$ and also a background field doublet $\cB_{a b}$ containing the two forms $B^{(2)}$ and $C^{(2)}$ 
and  using $\tilde{(\tilde{F})}=-F$, one can write the transformation of these field doublets as
\begin{eqnarray}
\cF_{a b}
 \rightarrow  (\Lambda^{-1})^T \cF_{a b}
\,\,\,\,\,\,\,;\,\,\,\,\,\,\,\,\,\cB_{a b}\labell{FB}
 \rightarrow  (\Lambda^{-1})^T \cB_{a b}.
\end{eqnarray}




Using the above transformations, one can find that the structures $\cF^T\cM\cF $ and $\cB^T\cM\cB $ are $SL(2,R)$ invariant structures.

Consider Lagrangian in terms of  two possible Lorentz invariants  $t=1/4F^{a b} F_{a b}$ and  $z=1/4F^{a b} \tilde{F}_{ab}$. At the presence of axion and dilaton couplings, it could be useful to express  the Lagrangian    in the following form which the contribution of axion coupling is separated.
\beqa
\cL=\cL^{'}+C_0z
\eeqa 
The antisymmetric tensor $G_{ab}$ then separates as $G_{ab}=G^{'}_{ab}-C_0\tilde{F}_{ab}$ where $G^{'}_{ab}$ could be an arbitrary nonlinear function of $F$ which is $G^{'}_{ab}=-2\frac{\prt \mathcal{L}^{'}}{\prt F^{ab}}$. 

By this consideration one gets the $SL(2,R)$ invariant structure in the following form:
\beqa
(\cF^T)_a{}^c\cM_0\cF_{bc}=e^{-\phi_0}\,\tilde{F}_a{}^c\,\tilde{F}_{bc}+ e^{\phi_0}\, G^{'}_{a}{}^c\, G^{'}_{bc},
\eeqa 
where the matrix $\cM_0$ is the matrix $\cM$ in which the dilaton and the R-R scalar are constant ($\phi_0, C_0$).
It is easy to check that the above structure is invariant under the linear transformations $G^{'}   \longrightarrow  \tilde{F}$ , $\tilde{F} \longrightarrow   -G^{'}$ and $e^{-\phi_0}  \longrightarrow  e^{\phi_0}$.

َWhen the dilaton and the R-R scalar are constant, one finds the following S-dual multiplet\cite{Garousi:1201}:
\beqa
(\tilde{\cF}^T)_a{}^c\cM_0\cB_{bc}&=&e^{-\phi_0}F_a{}^cB_{bc}-\tilde{F}_a{}^cC_{bc}^{(2)}-C_0\tilde{F}_a{}^cB_{bc}.
\eeqa
At the presence of two gauge fields with two derivatives, one can find the following  $SL(2,R)$ invariant structure
\beqa
\prt \cF^T\cM\prt \cF_{}&=&e^{-\phi_0} \prt \tilde{F} \prt \tilde{F}+ e^{\phi_0} \prt G^{'} \prt  G^{'}.\labell{FMF}
\eeqa
By considering the nonlinear expansion of $G'_{ab}$ (in any self dual theory) as $G^{'}_{ab}=e^{-\phi_0} F+e^{-2\phi_0} F^3+\cdots +e^{-N\phi_0}F^{2N-1}$,
 the nonlinear $SL(2,R)$ invariant structure corresponding to $2N$ gauge fields and  two derivatives could be found.
\beqa
\prt \cF^T\cM\prt \cF_{}&\sim& \prt F  \prt F+ F F \prt F  \prt F+ F F F F \prt F  \prt F+\cdots.
\eeqa
\section{S-dual amplitude of one gauge field, two scalar fields and one two-form}

The S-matrix elements of one gauge field, two scalar fields and one two-form vertex operator could be calculated from $D_3$-brane effective action. At first, we are going to find the corresponding amplitude in which the two-form is an antisymmetric $RR$ field $C^{(2)}$. The amplitude is given by the following Feynman rule:
\beqa
\cA(C_1^{(2)}, F_2, \Phi_3, \Phi_4)=V^a(C_1^{(2)},A)G_{ab}(A)V^b(A,F_2,\Phi_3, \Phi_4),\labell{A}
\eeqa
where $F_2$ is the polarization of the gauge boson, $\Phi_3, \Phi_4$ are the scalar fields and $A$ is the off-shell gauge field propagating between the two
vertices. The gauge boson propagator and the vertices  can be read from $D_3$-brane effective action
\beqa
G_{ab}(A)&\sim&-\frac{e^{\phi_0}\eta_{ab}}{k\inn k}\nonumber\\ 
 V^a(C_1^{(2)},A)&\sim&-k\inn(\tilde{C_1}^{(2)})^a \\\labell{ver}
V^b(A,F_2,\Phi_3, \Phi_4)&\sim& e^{-\phi_0}{\Phi_3}_{i}{\Phi_4}^{i}\bigg[k\inn k_3 k_4\inn F_2^a +k\inn F_2\inn k_4 k_3^a-\frac{1}{2} k_3\inn k_4 k\inn F_2^a\bigg]+(3\leftrightarrow 4)\nonumber
\eeqa
where $(\tilde{C}^{(2)})_{ab}=\frac{1}{2}\eps_{abcd}(C^{(2)})^{cd}$. From the conservation of momentum
along the brane we have $k=-k_1=k_2+k_3+k_4$. Considering the above vertices and propagator, one finds that  the S-matrix elements have not any contribution to the contact amplitude and  the amplitude \reef{A} then appears as a pole amplitude:
\beqa
\cA(C^{(2)}_1,F_2,\Phi _3,\Phi _4) &=&\frac{{\Phi _3}_{i}\Phi _4^{i}}{k_1\inn k_1}\bigg[  k_1\inn \tilde{C}^{(2)}_1\inn k_4  \,\, k_1\inn F_2\inn k_3- k_1\inn k_3  \,\, k_1\inn \tilde{C}^{(2)}_1\inn F_2 \inn k_4\labell{AC}\nonumber\\
&&\,\,\,\,\,\,\,\,\,\,\,\,\,\,\,\,+\frac{1}{2} k_3\inn k_4\,\, k_1\inn \tilde{C}^{(2)}_1 \inn F_2 \inn k_1\bigg]+(3\leftrightarrow 4).
\eeqa 
The amplitude dose not satisfy the Ward identity corresponding to the S-duality transformations, however is invariant under the Ward identity corresponding to abelian gauge symmetry and to the RR gauge symmetry.

The amplitude corresponding to antisymmetric B-field has massless pole as well as contact term.  The contact part of B-field amplitude can be found using the Feynman rule and considering the relevant higher derivative correction of D-brane action.
\beqa
\cA_{contact}\left(B_1,F_2,\Phi_3,\Phi_4\right)=e^{-\phi _0}{\Phi _3}_{i}\Phi _4^{i}\bigg[\frac{1}{4} k_3\inn k_4 Tr(B_1\inn F_2)- k_3\inn B_1\inn F_2\inn k_4\bigg]+(3\leftrightarrow 4).\labell{ABc}
\eeqa
Considering the propagator and vertex $V^b(A,F_2,\Phi_3, \Phi_4)$ in \reef{ver} and using the fourth order in the expansion of D-brane action to find the vertex operator $V^a(B_1,A)=e^{-\phi_0}k_1\inn B_1^a$, one finds the pole part of B-field amplitude as following:
\beqa
\cA_{pole}\left(B_1,F_2,\
\Phi _3,\Phi _4,\right) &=& V^a(B_1,A)G_{ab}(A)V^b( A,F_2,\Phi _3,\Phi _4)\nonumber\\
&=&-\frac{e^{-\Phi }{\Phi _3}_{i}\Phi _4^{i}}{k_1. k_1}\bigg[  k_1\inn B_1\inn k_4 k_1\inn F_2\inn k_3- k_1\inn k_3 k_1\inn B_1\inn F_2 \inn k_4\labell{ABp}\nonumber\\
&&\,\,\,\,\,\,\,\,\,\,\,\,\,\,\,\,\,\,\,\,\,\,\,\,\,\,\,\,\,+\frac{1}{2} k_3\inn k_4 k_1\inn B_1 \inn F_2 \inn k_1\bigg]+(3\leftrightarrow 4)
\eeqa
where we use the previously mentioned conservation of momentum.

Now, we are going to determine the transformation behavior of the amplitude \reef{AC} under the S-duality transformation  when the axion background dos not tack into account $C_0=0$. There are two different structures in amplitude \reef{AC} that transform under the S-duality as following:
\beqa
k_N\inn \tilde{C}^{(2)}_1 \inn F_2 \inn k_M & \longrightarrow  & e^{-\phi } k_N\inn \tilde{B}_1\inn \tilde{F}_2\inn k_M\nonumber\\
&=&e^{-\phi }\bigg(k_N\inn F_2\inn B_1\inn k_M-\frac{1}{2} k_N\inn k_M\Tr(B_1\inn F_2)\bigg),\nonumber\\
k_N\inn \tilde{C}^{(2)}_1\inn k_M\,\,\,\,\, k_P\inn F_2\inn k_Q & \longrightarrow & e^{-\phi} k_N\inn \tilde{B}_1\inn k_M\,\, k_P\inn \tilde{F}_2\inn k_Q\nonumber\\
 &=&e^{-\phi}\bigg(-\frac{1}{2} k_N\inn k_Q k_P\inn k_M \Tr(B_1\inn F_2)+\frac{1}{2} k_N\inn k_P k_Q\inn k_M \Tr(B_1\inn F_2)\nonumber\\
&&-k_N\inn F_2\inn k_M k_P\inn B_1\inn k_Q-k_Q\inn k_M k_N\inn F_2\inn B_1\inn k_P+k_P\inn k_M k_N\inn F_2\inn B_1\inn k_Q\nonumber\\
&&+k_N\inn k_Q k_M\inn F_2\inn B_1\inn k_P-k_N\inn k_P k_M\inn F_2\inn B_1\inn k_Q\bigg),\nonumber
\eeqa
where we use the following identity:
\beqa
\eps^{abcd}\eps^{efgh}=-\left|\begin{array}{cccc}
\eta^{ae}& \eta^{af}& \eta^{ag}&\eta^{ah}\cr  \
 \eta^{be}& \eta^{bf}& \eta^{bg}&\eta^{bh}\cr \
 \eta^{ce}& \eta^{cf}& \eta^{cg}&\eta^{ch}\cr \
\eta^{de}& \eta^{df}& \eta^{dg}&\eta^{dh}\\
\end{array}\right|.
 \labell{ee} 
\eeqa
Applying the above relations in the amplitude \reef{AC}, one obtains the amplitudes \reef{ABc} and \reef{ABp}. In fact the amplitude of one $RR$, two scalar and one gauge field transform to the amplitude of one $NSNS$ B-field, two scalar and one gauge field under the S-duality.
\beqa
\cA\left(C^{(2)}_1,F_2,\Phi _3,\Phi _4\right) & \longrightarrow &\cA_{contact}\left(B_1,F_2,\Phi _3,\Phi _4\right)+\cA_{pole}\left(B_1,F_2,\Phi _3,\Phi _4\right).
\eeqa
Therefore, to get the $NSNS$ coupling to brane, one can apply the S-duality on the pole amplitude contains $RR$ state. This is consistent with the statement that the $NSNS$ coupling is found by replacement $F \longrightarrow F+B$ in the D-brane effective action that has been expressed in literatures.

At the presence of the axion field, it is expected that one can find the combination of the amplitudes \reef{AC}, \reef{ABc} and \reef{ABp} in terms of $SL(2,R)$ invariant structures. To do this, we consider all contraction of the $SL(2,R)$ invariant structure $\tilde{\cF}^T\cM\cB$ with four, two and zero momenta as following:
\beqa
{({\tilde{\cF}_N}^{T})}^{ab}\cM \cB_{M}^{cd}\,\,{k_1}_a {k_2}_b {k_3}_c {k_4}_d&=&e^{-\phi }k_1\inn F_N\inn k_2 k_4\inn B_M\inn k_3 +C_0 e^{-\phi }k_1\inn F_N\inn k_2 k_4\inn \tilde{B}_M\inn k_3 \nonumber\\
&&-\frac{1}{2} k_4\inn k_2 k_1\inn k_3 \Tr(F_N\inn \tilde{C}_M)+\frac{1}{2} k_1\inn k_4 k_2\inn k_3 \Tr(F_N\inn \tilde{C}_M)\nonumber\\
&&+k_3\inn F_N\inn k_4 k_1\inn\tilde{C}_M \inn k_2-k_2\inn k_3 k_4\inn F_N\inn\tilde{C}_M\inn k_1\nonumber\\
&&+k_1\inn k_3 k_4\inn F_N\inn\tilde{C}_M\inn k_2+k_4\inn k_2 k_3\inn F_N\inn\tilde{C}_M\inn k_1\nonumber\\
&&-k_1\inn k_4 k_3\inn F_N\inn\tilde{C}_M\inn k_2,\\
k_3\inn \tilde{\cF}_N^T\cM\cB_M\inn k_4&=&-e^{-\phi } k_3\inn F_N\inn B_M\inn k_4-C_0 e^{-\phi } k_3\inn F_N\inn \tilde{B}_M\inn k_4\nonumber\\
&&+\frac{1}{2} k_3\inn k_4 \Tr(F_N\inn \tilde{C}_M)-k_3\inn \tilde{C}_M\inn F_N\inn k_4,\nonumber\\
\Tr(\tilde{\cF}_N^T\cM\cB_M)&=&-e^{-\phi } \Tr(F_N\inn B_M)-C_0e^{-\phi } \Tr(F_N\inn B_M)+\Tr(F_N\inn \tilde{C}_M).\nonumber
\eeqa
Considering the above relations, one can find the following combination of above $SL(2,R)$ invariant structures contains the amplitude \reef{AC} as well as the amplitude \reef{ABc} $+$ \reef{ABp}.
\beqa
\widetilde{\cA}_{}&=&-\frac{{\Phi _3}_{i}\Phi _4^{i}}{k_1\inn k_1}\bigg[\frac{1}{2}k_3\inn k_4 k_1\inn \tilde{\cF}_2^T\cM\cB_1\inn k_1- k_1\inn k_4 k_3\inn \tilde{\cF}_2^T\cM\cB_1\inn k_1-k_1\inn \tilde{\cF}_2^T\inn k_3\cM k_4\inn\cB_1\inn k_1\bigg]\nonumber\\
&&+{\Phi _3}_{i}\Phi _4^{i}\bigg[k_3\inn \tilde{\cF}_2^T\cM\cB_1\inn k_4-\frac{1}{4}k_3\inn k_4 Tr(\tilde{\cF}_2^T\cM\cB_1)\bigg]+(3\leftrightarrow 4).\labell{ASL2R}
\eeqa
It is clear that this $SL(2,R)$ invariant amplitude (that is manifestly invariant under the $SL(2,R)$ transformation) also contains the amplitude $\cA\left(C_0;B_1,F_2,\Phi _3,\Phi _4\right)$. In fact, to make a $SL(2,R)$ invariant amplitude from amplitudes \reef{AC} and \reef{ABc} $+$ \reef{ABp}, the amplitude corresponding to axion field should be take into account. On the other hand, we can confirm the above $SL(2,R)$ invariant amplitude by deriving  the axion amplitude from the explicit calculation and the Feynman rule that is resulted as following pole amplitude:
\beqa
\cA\left(C_0;B_1,F_2,\Phi _3,\Phi _4\right) &=&-\frac{C_0e^{-\phi }{\Phi _3}_{i}\Phi _4^{i}}{k_1\inn k_1}\bigg[  k_1\inn \tilde{B}_1\inn k_4 k_1\inn F_2\inn k_3- k_1\inn k_3 k_1\inn \tilde{B}_1\inn F_2 \inn k_4\labell{AC0}\nonumber\\
&&\,\,\,\,\,\,\,\,\,\,\,\,\,\,\,\,+\frac{1}{2} k_3\inn k_4 k_1\inn \tilde{B}_1 \inn F_2 \inn k_1\bigg]+(3\leftrightarrow 4)
\eeqa
This is exactly equal to the axion amplitude that lies in  \reef{ASL2R}.

\section{S-dual amplitude of three gauge fields and one two-form}
It has been shown that the  amplitude of three gauge fields and one two-form in which the two-form is $RR$ field $C^{(2)}$ or $NSNS$ B-field could be appeared in terms of $SL(2,R)$ invariant structures in the presence of the background dilaton and R-R scalar fields \cite{hosein:1304}. We are going to show that these two amplitudes transform to each other under the S-duality transformation when the axion field dose not tack into account. We strart with the amplitude of three gage fields and one $RR$ field $C^{(2)}$ that is founded by explicit calculation\cite{hosein:1304,ehsan}. 
\beqa
\cA\left({C_1}^{(2)}, F_2,F_3,F_4\right) &=&\frac{1}{4}e^{- \phi _0}\frac {k_1\inn F_2\inn \tilde{C}^{(2)}_1\inn k_1}{k_1\inn k_1}Tr(F_3\inn F_4)\nonumber\\
&&-e^{-\phi _0}\frac {k_1\inn F_2\inn F_3\inn F_4\inn \tilde{C}^{(2)}_1\inn k_1}{k_1\inn k_1}+P(2,3,4),\labell{ACFFF}
\eeqa
where $P(2, 3, 4)$ stands for the other permutations of $2, 3, 4$. This amplitude contains two independent structures. We apply S-duality transformation to these two structures separately.
\beqa
e^{- \phi _0} k_1\inn F_2\inn F_3\inn F_4\inn \tilde{C}^{(2)}\inn k_1 && \mathop{\longrightarrow}^{S} \,\,\,\, \frac{e^{-2 \phi _0}}{16}k_{1a} \epsilon ^{abkl}\epsilon _{bcnm}\epsilon ^{cdgh}\epsilon _{deij}F_{2kl}{F_3}^{nm}F_{4gh}B^{ij}{k_1}^e=I_{234},\nonumber\\
e^{- \phi _0}k_1\inn F_2\inn \tilde{C}^{(2)}\inn k_1 Tr(F_3\inn F_4)&&\mathop{\longrightarrow}^{S}\,\,\,\,\frac{e^{-2 \phi _0}}{16}k_{1a} \epsilon ^{abkl}\epsilon _{bcnm}\epsilon ^{degh}\epsilon _{deij}F_{2kl}{F_3}_{gh}{F_4}^{ij}B^{nm}{k_1}^e=J_{234}.\nonumber
\eeqa
 Because the invariance of the trace term in the original amplitude under the gauge fields permutations, one can easily conclude $J_{234}=J_{243}$, $J_{423}=J_{432}$ and $J_{324}=J_{342}$  in the dual amplitude. Therefore, the gauge field permutations can produce six different terms of $I$-type and three different terms of $J$-type that contribute in the dual amplitude.  

The calculation in this case is more complicated than the case in the previous section due to the presence of four Levi-Civita tensors. To rewrite this terms in terms of various contractions of gauge fields, one has to use the identity \reef{ee}. Here, one encounters with different choices. In fact, there are three ways to pairs the Levi-Civita tensors: $\cE=\big( \epsilon ^{abkl}\epsilon _{bcnm}\big)\big( \epsilon ^{degh}\epsilon _{deij}\big)$, $\cE'=\big( \epsilon ^{abkl}\epsilon _{deij}\big)\big( \epsilon ^{cdgh}\epsilon _{bcnm}\big) $ and $\cE''=\big( \epsilon ^{abkl}\epsilon ^{cdgh}\big)\big(\epsilon _{bcnm}\epsilon _{deij}\big)$. Using the standard identity \reef{ee} for the above first two parings in I-type terms, we have:
\beqa
I_{NMP}(\cE)&=&e^{-2 \phi _0}\bigg[-k_1\inn F_{N}\inn B\inn F_{P}\inn F_{M}\inn k_1-k_1\inn F_{N}\inn F_{M}\inn  F_{P}\inn B\inn k_1- k_1\inn F_3\inn F_{M}\inn F_{N}\inn B\inn k_1\nonumber\\
&&\,\,\,\,\,\,\,\,\,\,\,\,\,\,+\frac{1}{2} k_1\inn F_{P}\inn F_{M}\inn k_1 Tr(F_{N}\inn B)+\frac{1}{2} k_1\inn F_{N}\inn B\inn k_1 Tr(F_{P}\inn F_{M})\nonumber\\
&&\,\,\,\,\,\,\,\,\,\,\,\,\,\,-\frac{1}{4} k_1\inn k_1 Tr(F_{N} \inn B) Tr(F_{P}\inn F_{M})+ k_1\inn k_1 Tr(F_{N}\inn F_{M}\inn F_{P} \inn B)\bigg]\nonumber\\
I_{NMP}(\cE')&=&e^{-2 \phi _0}\bigg[ k_1\inn  F_{P}\inn B\inn F_{N}\inn F_{M}\inn k_1-\frac{1}{2} k_1\inn  F_{P}\inn B\inn k_1 Tr(F_{N}\inn F_{M})\nonumber\\
&&\,\,\,\,\,\,\,\,\,\,\,\,\,\,\,\,\,\,\,\,\,\,\,-\frac{1}{2} k_1\inn F_{N}\inn F_{M}\inn k_1 Tr( F_{P}\inn B)+\frac{1}{4} k_1\inn k_1 Tr(F_{N} \inn F_{M})Tr(F_{P}\inn B)\bigg]\labell{iee'}
\eeqa
and for the first paring in the J-type term, we have 
\beqa
J_{NMP}(\cE)&=&e^{-2 \phi _0}\bigg[- k_1\inn F_{N}\inn B\inn k_1 Tr(F_{M}\inn F_{P})+\frac{1}{2} k_1\inn k_1 Tr(F_{N} \inn B) Tr(F_{M}\inn F_{P})\bigg]\labell{je}
\eeqa
where $N,M,P=2,3,4$ and $N\neq M\neq P$. 

Our calculations show that one can get to the amplitude of $\cA\left({B_1}^{(2)}, F_2,F_3,F_4\right)$ only by selecting the specific parings of  Levi-Civita tensors as following:
\beqa
&&\cA\left({C_1}^{(2)},F_2,F_3,F_4\right)  \mathop{\longrightarrow}^{S}\frac {I_{234}(\cE')+I_{324}(\cE')+I_{423}(\cE')+I_{243}(\cE)+I_{342}(\cE)+I_{432}(\cE)}{k_1\inn k_1}\nonumber\\
&&\,\,\,\,\,\,\,\,\,\,\,\,\,\,\,\,\,\,\,\,\,\,\,\,\,\,\,\,\,\,\,\,\,\,\,\,\,\,\,\,\,\,\,\,\,\,\,\,\,\,\,\,\,\,\,\,\,\,\,-\frac{1}{2}\frac {J_{234}(\cE)+J_{324}(\cE)+J_{423}(\cE)}{k_1 \inn k_1}.\labell{aa}
\eeqa
Replacing \reef{iee'} and \reef{je} in the above relation and using the conservation of momentum, one finds the following expression, after a straightforward calculation.
\beqa
&&\frac{1}{2}e^{-2 \phi _0} \left( \Tr(B_1\inn F_2\inn F_3\inn F_4)-\frac{1}{4}  \Tr(B_1\inn F_2)\Tr(F_3\inn F_4)\right)\nonumber\\
&&-\frac{1}{4}e^{-2 \phi _0}\frac {k_1\inn F_2\inn B_1\inn k _1}{k_1\inn k_1}Tr(F_3 \inn F_4)+e^{-2 \phi _0}\frac {k_1\inn F_2\inn F_3\inn  F_4\inn B_1\inn k _1}{k_1\inn k_1}+P(2,3,4).\labell{BFFF}
\eeqa
As we have mentioned in \reef{AB}, this is exactly equal to the amplitude of one $B$-field and three gauge fields containing pole term as well as contact term that calculated in \cite{hosein:1304}.

\section{S-dual effective action with four derivative corrections}

It has been shown that the one-loop correction to the $D_3$-brane Abelian Born-Infeld action \reef{LF4} containing four gauge fields with four derivative proportional to $(s^2+t^2+u^2)F^4$ where $s,t$ and $u$ are the Mandelstam variables \cite{Shmakova}. This result has been found from supersymmetry fixing \cite{De} and string Disk-level scattering amplitude calculation\cite{Hashimoto}. The $D_3$-brane effective action at the order of $(\partial F)^4$ is invariant under the $SL(2,R)$ duality. From the Lagrangian viewpoint, this action saturates the NGZ identity\cite{Chemissany}.

 We are going to find this action in terms of some $SL(2,R)$ invariant structures. To do this, one should consider all contractions of two nonlinear structures $W^{abcdefg}W_{ghnmpq}$, which any of them contains two gauge field derivative terms as following
\beqa
W^{abcdef}=\partial^{a} {(\cF^T)}^{bc} \cM  \,\,\,\, \partial^{d}{\cF}^{ef}.
\eeqa
Using "xAct" \cite{Nutma}, one finds the following 33 different $WW$ contractions:
\beqa
&& w_1 \
W_{c}{}^{efcab} W_{def}{}^{d}{}_{ab} + w_2 W_{a}{}^{ab}{}_{c}{}^{ef} \
W_{def}{}^{d}{}_{b}{}^{c} + w_3 W_{bef}{}^{cab} W_{dc}{}^{fe}{}_{a}{}^{d} +  w_4 W_{bc}{}^{fcab} \
W_{def}{}^{e}{}_{a}{}^{d}  \nonumber\\
&& + w_5 W_{cdf}{}^{cab} \
W_{eb}{}^{fe}{}_{a}{}^{d} +w_6 \
W_{a}{}^{ab}{}_{cb}{}^{c} W_{d}{}^{de}{}_{fe}{}^{f} + w_7 W_{a}{}^{ab}{}_{cd}{}^{e} \
W^{d}{}_{b}{}^{c}{}_{fe}{}^{f} + w_8 W_{a}{}^{ab}{}_{dc}{}^{e} \
W^{d}{}_{b}{}^{c}{}_{fe}{}^{f}  \nonumber\\
&& + w_9 W_{a}{}^{ab}{}_{b}{}^{cd} \
W^{e}{}_{cdfe}{}^{f} + w_{10} W_{a}{}^{abd}{}_{b}{}^{c} \
W^{e}{}_{cdfe}{}^{f}  
+ w_{11} W^{cabe}{}_{a}{}^{d} W_{fce}{}^{f}{}_{bd}+ w_{12} W_{cdf}{}^{cab} \
W^{e}{}_{a}{}^{df}{}_{be} \nonumber\\
&&  + w_{13} W^{cab}{}_{dcf} \
W^{e}{}_{a}{}^{df}{}_{be} + w_{14} W^{cabe}{}_{a}{}^{d} W_{fcd}{}^{f}{}_{be} \
+ w_{15} W^{cab}{}_{def} \
W^{d}{}_{ac}{}^{f}{}_{b}{}^{e} + w_{16} W_{ca}{}^{dcab} \
W_{edf}{}^{f}{}_{b}{}^{e}  \nonumber\\
&& + w_{17} W^{cabd}{}_{ac} \
W_{edf}{}^{f}{}_{b}{}^{e} + w_{18} W_{ca}{}^{dcab} \
W_{fde}{}^{f}{}_{b}{}^{e} + w_{19} W^{cabd}{}_{ac} \
W_{fde}{}^{f}{}_{b}{}^{e}+ w_{20} W_{a}{}^{ab}{}_{b}{}^{cd} \
W_{def}{}^{f}{}_{c}{}^{e}  \nonumber\\
&& + w_{21} W^{cab}{}_{def} \
W^{d}{}_{ab}{}^{f}{}_{c}{}^{e} + w_{22} W_{a}{}^{ab}{}_{def} \
W^{d}{}_{b}{}^{cf}{}_{c}{}^{e} + w_{23} W^{cabd}{}_{ab} \
W_{edf}{}^{f}{}_{c}{}^{e} +  w_{24} W_{a}{}^{abd}{}_{b}{}^{c} \
W_{edf}{}^{f}{}_{c}{}^{e}  \nonumber\\
&&  + w_{25} W^{cabd}{}_{ab} \
W_{fde}{}^{f}{}_{c}{}^{e}+ w_{26} W_{a}{}^{abd}{}_{b}{}^{c} \
W_{fde}{}^{f}{}_{c}{}^{e}+w_{27} W_{bef}{}^{cab} \
W^{d}{}_{ac}{}^{f}{}_{d}{}^{e} + w_{28} W_{a}{}^{ab}{}_{cef} \
W^{d}{}_{b}{}^{cf}{}_{d}{}^{e}  \nonumber\\
&& + w_{29} W_{a}{}^{ab}{}_{cb}{}^{c} \
W_{edf}{}^{fde} + w_{30} W_{bac}{}^{cab} W_{edf}{}^{fde} + w_{31} \
W_{cab}{}^{cab} W_{edf}{}^{fde}   + w_{32} W_{a}{}^{ab}{}_{cb}{}^{c} \
W_{fde}{}^{fde}  \nonumber\\
&&+ w_{33} W_{cab}{}^{cab} W_{fde}{}^{fde},\labell{WW}
\eeqa
where $w_i$ are some unknown coefficients that would be fiexed by imposing some physical identity.

 Using the $SL(2,R)$ invariant structure \reef{FMF} at the level of four gauge fields and four derivatives, one finds that the $WW$ structures  appear in terms of the gauge field contents $(\partial F)^4, (\partial F)^2(\partial \tilde{F})^2$ and $(\partial \tilde{F})^4$.  In order to make the action \reef{LF4} in terms of the above $WW$ structures, we should find the contribution of dual gauge field $\tilde{F}$ in the last two field contents in terms of gauge field $F$ by using the identity \reef{ee}. For the third field content, we encounter with four Levi-Civita tensors and with different choices to pair these tensors as was mentioned in the previous section:  $\cE, \cE'$ and $\cE''$. Any of these parings of Levi-Civita tensors led to different results in which the gauge fields appear in different kinds of contraction terms $(\partial F)^4$. But one can show that these results are equal when the identities presenting in the appendix have been tack into account.

After finding the above $WW$ structures in terms of gauge field contraction terms $(\partial F)^4$ and using the identities in the appendix properly, we find the following 9 equations between the constants:
\beqa
w_{15}&=&  -\frac{3}{2}+2( w_{10} +  w_{17} -  w_{20} -  w_{24})  - 3  (w_{29 }-  w_{30}-  w_{6})+  w_{22} +  w_{8}, \nonumber\\
 w_{13}&=&\frac{2}{3}(  w_{10} +  w_{15} -  w_{20} -  w_{22}  -  w_{8})+\frac{1}{3} ( -  w_{17}+  w_{24}), \nonumber\\
w_{21}&=&\frac{2}{3}(w_{10} +  w_{17 } - w_{20} -  w_{22} -  w_{24})+\frac{1}{2}(w_{12} -   w_{27} -   w_{7})+ \frac{1}{6}  w_{15}
 -\frac{7}{6}w_{8} + w_{16}+ w_{2}, \nonumber\\
w_{26 }&=&-\frac{1}{3}(w_{8}-  w_{15}) +\frac{2}{3}(-  w_{10} +   w_{20}+   w_{22} )+\frac{8}{3}(-   w_{17}+   w_{24 })+6( w_{29 }-  w_{30} - w_{6})\nonumber\\
&&\,\,+ w_{14} -  w_{16}  +  w_{19} +  w_{27 } -  w_{7 }, \nonumber\\
 w_{28}&=&\frac{4}{3}(- w_{10}-  w_{17} + w_{20}+  w_{24})+\frac{1}{3}(w_{8} -  w_{15 } +  w_{22} )+4(  w_{29} -  w_{30} - w_{6})
+  w_{27}  - w_{7}, \nonumber\\
 w_{3}&=&\frac{4}{3}( w_{10}- w_{20} ) +\frac{1}{3} ( w_{15} +  w_{17} -  w_{22} -  w_{24} -w_{8}) -  w_{29} + w_{30} + w_{6}, \nonumber\\
 w_{4}&=&  w_{17} -  w_{24} -  w_{29} +  w_{30} +  w_{6}, \nonumber\\
 w_{5}&=& \frac{2}{3}( w_{10}  -   w_{20}-   w_{8})+\frac{1}{2}( w_{14 }+   w_{29}-   w_{30} -   w_{6} -   w_{7} )+ \frac{1}{6}(w_{15 }+  w_{17} -  w_{22} -  w_{24})\nonumber\\
&&\,-2 w_{1}  - w_{18} + w_{2} \nonumber\\
 w_{9}&=&\frac{1}{6} (- w_{10} -  w_{15} -  w_{17 }+  w_{20} +  w_{22} +  w_{24} +  w_{8}).\nonumber
\eeqa

The nonzero coefficients could be fixed as follows by comparing the result with the action \reef{LF4} :
\beqa
\ w_3= w_4=w_5= w_6= \frac{1}{2},\,\,\,\,\,\,\,\,\,\,\,\, w_{7}= - \frac{3}{2},\,\,\,\,\,\,\,\,\,\,\,\,\, w_{27}= \frac{3}{2} ,\,\,\,\,\,\,\,\,\,\,\,\,\,\, w_{28} = 1. \nonumber
\eeqa
By substituting these nonzero coefficients in \reef{WW}, we can find the $SL(2,R)$ invariant form of the action \reef{LF4}.
\beqa
  L_{(\partial F{})^{4}}^{S}&=&\frac{1}{2} W_{bef}{}^{cab} W_{dc}{}^{fe}{}_{a}{}^{d} + \frac{1}{2} W_{a}{}^{ab}{}_{cb}{}^{c} W_{d}{}^{de}{}_{fe}{}^{f} + \frac{1}{2} W_{bc}{}^{fcab} W_{def}{}^{e}{}_{a}{}^{d}+ \frac{1}{2} W_{cdf}{}^{cab} W_{eb}{}^{fe}{}_{a}{}^{d}  \nonumber\\
&& -  \frac{3}{2} W_{a}{}^{ab}{}_{cd}{}^{e} W^{d}{}_{b}{}^{c}{}_{fe}{}^{f}+\frac{3}{2} W_{bef}{}^{cab} W^{d}{}_{ac}{}^{f}{}_{d}{}^{e} + W_{a}{}^{ab}{}_{cef} W^{d}{}_{b}{}^{cf}{}_{d}{}^{e}. \labell{LS}
\eeqa

In order to reach the action at the order of six gauge fields and four derivatives, one should consider the expansion of the field $G^{'}$ up to the level of three gauge field $F^3$. So the expansion of $\partial G^{'}$ up to this level, could be found as follows:
\beqa
\partial_{a}{G^{'}}_{bc}&=&e^{-\phi_0} \partial_{a}F_{bc} +e^{-2\phi_0}\bigg( \frac{1}{4} F_{de} F^{de} \
\partial_{a}F_{bc} + F_{c}{}^{d} F_{d}{}^{e} \
\partial_{a}F_{be} -   F_{b}{}^{d} F_{d}{}^{e} \
\partial_{a}F_{ce}\nonumber\\
&& -   F_{b}{}^{d} F_{c}{}^{e} \
\partial_{a}F_{de} + \frac{1}{2} F_{bc} F^{de} \
\partial_{a}F_{de}\bigg)+\cdots\nonumber
\eeqa
  where dots refer to the terms at higher order of gauge field.
  
   Inserting the above expansion into the nonzero invariant structurs $WW$ in \reef{LS}, one can find the action containing six gauge fields with four derivatives that is compatible with S-duality. This action could be appear in the simple form when one consider the total derivative terms. Actualy, at the level of six gauge fields and four derivatives, there are 584 contractions with structure $FFDFDFDFDF$. To find the total derivative terms, we note that there are 724 total derivative terms with structure $D[FFFDFDFDF]$. Using their coefficients to eliminate the terms with structure $FFFDDFDFDF$, one finds 483 constraint equations. Applying these constraints,  the total derivative terms with structure $FFDFDFDFDF$ could be found. Considering these terms and  using the the on-shell condition $\partial_{a}F^{ab}=0$
, one can find the action corresponding to six gauge fields with four derivatives as following simple form.
\beqa
 L_{F^2(\partial F{})^{4}}^S\hspace{-.25cm}&\sim &   \frac{3}{2} F^{ab} F^{cd} \partial_{c}F_{e}{}^{f} 
\partial_{d}F^{hi} \partial^{e}F_{ab} \partial_{f}F_{hi} + 
\frac{1}{4} F^{ab} F^{cd} \partial_{d}F_{c}{}^{f} 
\partial_{e}F^{hi} \partial^{e}F_{ab} \partial_{f}F_{hi} \nonumber\\
&&- 3 
F^{ab} F^{cd} \partial_{b}F_{e}{}^{f} \partial_{d}F^{hi} 
\partial^{e}F_{ac} \partial_{f}F_{hi} -  \frac{1}{2} F^{ab} 
F^{cd} \partial_{d}F_{b}{}^{f} \partial_{e}F^{hi} 
\partial^{e}F_{ac} \partial_{f}F_{hi}\nonumber\\
&& + 3 F_{a}{}^{c} F^{ab} 
\partial_{c}F^{hi} \partial_{d}F_{e}{}^{f} \partial^{e}F_{b}{}^{d} 
\partial_{f}F_{hi} - 3 F_{a}{}^{c} F^{ab} \partial_{c}F_{e}{}^{f} 
\partial_{d}F^{hi} \partial^{e}F_{b}{}^{d} \partial_{f}F_{hi} \nonumber\\
&&+ 
\frac{1}{2} F_{a}{}^{c} F^{ab} \partial_{c}F_{d}{}^{f} 
\partial_{e}F^{hi} \partial^{e}F_{b}{}^{d} \partial_{f}F_{hi} -  
\frac{1}{2} F^{ab} F^{cd} \partial_{d}F^{hi} \partial^{e}F_{ac} 
\partial_{f}F_{hi} \partial^{f}F_{be} \nonumber\\
&&+ \frac{1}{4} F^{ab} F^{cd} 
\partial_{d}F^{hi} \partial^{e}F_{ab} \partial_{f}F_{hi} 
\partial^{f}F_{ce} -  F_{a}{}^{c} F^{ab} \partial_{d}F^{hi} 
\partial^{e}F_{b}{}^{d} \partial_{f}F_{hi} \partial^{f}F_{ce}\nonumber\\
&& + 
\frac{1}{2} F_{a}{}^{c} F^{ab} \partial_{c}F^{hi} 
\partial^{e}F_{b}{}^{d} \partial_{f}F_{hi} \partial^{f}F_{de} + 
F^{ab} F^{cd} \partial_{b}F_{ac} \partial_{e}F^{hi} 
\partial_{f}F_{hi} \partial^{f}F_{d}{}^{e}\nonumber\\
&& -  \frac{1}{2} F^{ab} 
F^{cd} \partial_{c}F_{ab} \partial_{e}F^{hi} \partial_{f}F_{hi} 
\partial^{f}F_{d}{}^{e} + F_{a}{}^{c} F^{ab} 
\partial_{c}F_{b}{}^{d} \partial_{e}F^{hi} \partial_{f}F_{hi} 
\partial^{f}F_{d}{}^{e}\nonumber\\
&& + 3 F^{ab} F^{cd} \partial_{d}F_{hi} 
\partial_{e}F_{f}{}^{i} \partial^{e}F_{ac} \partial^{h}F_{b}{}^{f} 
+ 3 F^{ab} F^{cd} \partial_{d}F_{fi} \partial^{e}F_{ac} 
\partial_{h}F_{e}{}^{i} \partial^{h}F_{b}{}^{f}\nonumber\\
&& - 2 F^{ab} F^{cd} 
\partial_{d}F_{e}{}^{i} \partial^{e}F_{ac} \partial_{h}F_{fi} 
\partial^{h}F_{b}{}^{f} -  \frac{3}{2} F^{ab} F^{cd} 
\partial_{d}F_{hi} \partial_{e}F_{f}{}^{i} \partial^{e}F_{ab} 
\partial^{h}F_{c}{}^{f} \nonumber\\
&&+ 3 F_{a}{}^{c} F^{ab} \partial_{e}F_{hi} 
\partial^{e}F_{b}{}^{d} \partial_{f}F_{d}{}^{i} 
\partial^{h}F_{c}{}^{f} - 2 F_{a}{}^{c} F^{ab} 
\partial_{e}F_{d}{}^{i} \partial^{e}F_{b}{}^{d} \partial_{f}F_{hi} 
\partial^{h}F_{c}{}^{f}\nonumber\\
&& + 3 F_{a}{}^{c} F^{ab} 
\partial^{e}F_{b}{}^{d} \partial_{f}F_{ei} \partial_{h}F_{d}{}^{i} 
\partial^{h}F_{c}{}^{f} -  \frac{3}{2} F^{ab} F^{cd} 
\partial_{d}F_{fi} \partial^{e}F_{ab} \partial_{h}F_{e}{}^{i} 
\partial^{h}F_{c}{}^{f} \nonumber\\
&&+ F^{ab} F^{cd} \partial_{d}F_{e}{}^{i} 
\partial^{e}F_{ab} \partial_{h}F_{fi} \partial^{h}F_{c}{}^{f} - 4 
F_{a}{}^{c} F^{ab} \partial_{e}F_{d}{}^{i} \partial^{e}F_{b}{}^{d} 
\partial_{h}F_{fi} \partial^{h}F_{c}{}^{f}\nonumber\\
&& - 3 F_{a}{}^{c} F^{ab} 
\partial_{c}F_{hi} \partial_{e}F_{f}{}^{i} \partial^{e}F_{b}{}^{d} 
\partial^{h}F_{d}{}^{f} - 3 F_{a}{}^{c} F^{ab} \partial_{c}F_{fi} 
\partial^{e}F_{b}{}^{d} \partial_{h}F_{e}{}^{i} 
\partial^{h}F_{d}{}^{f} \nonumber\\
&&+ 2 F_{a}{}^{c} F^{ab} 
\partial_{c}F_{e}{}^{i} \partial^{e}F_{b}{}^{d} \partial_{h}F_{fi} 
\partial^{h}F_{d}{}^{f} + 2 F^{ab} F^{cd} \partial_{c}F_{f}{}^{i} 
\partial_{d}F_{hi} \partial^{e}F_{ab} \partial^{h}F_{e}{}^{f}\nonumber\\
&& - 4 
F^{ab} F^{cd} \partial_{b}F_{f}{}^{i} \partial_{d}F_{hi} 
\partial^{e}F_{ac} \partial^{h}F_{e}{}^{f} + 4 F_{a}{}^{c} F^{ab} 
\partial_{c}F_{hi} \partial_{d}F_{f}{}^{i} \partial^{e}F_{b}{}^{d} 
\partial^{h}F_{e}{}^{f} \nonumber\\
&&- 4 F_{a}{}^{c} F^{ab} 
\partial_{c}F_{f}{}^{i} \partial_{d}F_{hi} \partial^{e}F_{b}{}^{d} 
\partial^{h}F_{e}{}^{f} -  \frac{3}{2} F_{ab} F^{ab} 
\partial_{e}F_{fi} \partial^{e}F^{cd} \partial^{h}F_{c}{}^{f} 
\partial^{i}F_{dh}\nonumber\\
&& + 2 F_{ab} F^{ab} \partial^{e}F^{cd} 
\partial_{f}F_{ei} \partial^{h}F_{c}{}^{f} \partial^{i}F_{dh} + 2 
F_{a}{}^{c} F^{ab} \partial_{e}F_{c}{}^{f} \partial^{e}F_{b}{}^{d} 
\partial_{f}F_{hi} \partial^{i}F_{d}{}^{h}\nonumber\\
&& + \frac{1}{2} F_{ab} 
F^{ab} \partial_{e}F_{c}{}^{f} \partial^{e}F^{cd} 
\partial_{h}F_{fi} \partial^{i}F_{d}{}^{h} + 2 F^{ab} F^{cd} 
\partial_{e}F_{c}{}^{f} \partial^{e}F_{ab} \partial_{i}F_{fh} 
\partial^{i}F_{d}{}^{h} \nonumber\\
&&- 4 F^{ab} F^{cd} \partial_{e}F_{b}{}^{f} 
\partial^{e}F_{ac} \partial_{i}F_{fh} \partial^{i}F_{d}{}^{h} - 4 
F_{a}{}^{c} F^{ab} \partial_{e}F_{c}{}^{f} \partial^{e}F_{b}{}^{d} 
\partial_{i}F_{fh} \partial^{i}F_{d}{}^{h} \nonumber\\
&&+ F_{ab} F^{ab} 
\partial_{e}F_{c}{}^{f} \partial^{e}F^{cd} \partial_{i}F_{fh} 
\partial^{i}F_{d}{}^{h} - 2 F_{a}{}^{c} F^{ab} 
\partial^{e}F_{b}{}^{d} \partial^{f}F_{ce} \partial_{i}F_{fh} 
\partial^{i}F_{d}{}^{h} \nonumber\\
&&+ F_{ab} F^{ab} \partial^{e}F^{cd} 
\partial^{f}F_{ce} \partial_{i}F_{fh} \partial^{i}F_{d}{}^{h} + 3 
F^{ab} F^{cd} \partial_{b}F_{ac} \partial_{d}F^{ef} 
\partial_{f}F_{hi} \partial^{i}F_{e}{}^{h}\nonumber\\
&& -  \frac{3}{2} F^{ab} 
F^{cd} \partial_{c}F_{ab} \partial_{d}F^{ef} \partial_{f}F_{hi} 
\partial^{i}F_{e}{}^{h} + 3 F_{a}{}^{c} F^{ab} 
\partial_{c}F_{b}{}^{d} \partial_{d}F^{ef} \partial_{f}F_{hi} 
\partial^{i}F_{e}{}^{h} \nonumber\\
&&-  \frac{1}{4} F_{ab} F^{ab} 
\partial^{e}F^{cd} \partial_{f}F_{hi} \partial^{f}F_{cd} 
\partial^{i}F_{e}{}^{h} + 3 F^{ab} F^{cd} \partial_{b}F_{ac} 
\partial_{f}F_{hi} \partial^{f}F_{d}{}^{e} \partial^{i}F_{e}{}^{h} \nonumber\\
&&
-  \frac{3}{2} F^{ab} F^{cd} \partial_{c}F_{ab} 
\partial_{f}F_{hi} \partial^{f}F_{d}{}^{e} \partial^{i}F_{e}{}^{h} 
+ 3 F_{a}{}^{c} F^{ab} \partial_{c}F_{b}{}^{d} \partial_{f}F_{hi} 
\partial^{f}F_{d}{}^{e} \partial^{i}F_{e}{}^{h}\nonumber\\
&& + \frac{3}{4} 
F_{ab} F^{ab} \partial^{e}F^{cd} \partial^{f}F_{cd} 
\partial_{h}F_{fi} \partial^{i}F_{e}{}^{h} + F^{ab} F^{cd} 
\partial_{d}F_{c}{}^{f} \partial^{e}F_{ab} \partial_{i}F_{fh} 
\partial^{i}F_{e}{}^{h} \nonumber\\
&&- 2 F^{ab} F^{cd} \partial_{d}F_{b}{}^{f} 
\partial^{e}F_{ac} \partial_{i}F_{fh} \partial^{i}F_{e}{}^{h} + 2 
F_{a}{}^{c} F^{ab} \partial_{c}F_{d}{}^{f} \partial^{e}F_{b}{}^{d} 
\partial_{i}F_{fh} \partial^{i}F_{e}{}^{h}\nonumber\\
&& + F^{ab} F^{cd} 
\partial_{b}F_{ac} \partial^{f}F_{d}{}^{e} \partial_{i}F_{fh} 
\partial^{i}F_{e}{}^{h} -  \frac{1}{2} F^{ab} F^{cd} 
\partial_{c}F_{ab} \partial^{f}F_{d}{}^{e} \partial_{i}F_{fh} 
\partial^{i}F_{e}{}^{h} \nonumber\\
&&+ F_{a}{}^{c} F^{ab} 
\partial_{c}F_{b}{}^{d} \partial^{f}F_{d}{}^{e} \partial_{i}F_{fh} 
\partial^{i}F_{e}{}^{h} -  F^{ab} F^{cd} \partial_{d}F_{hi} 
\partial_{e}F_{c}{}^{f} \partial^{e}F_{ab} \partial^{i}F_{f}{}^{h} \nonumber\\
&&
+ 2 F^{ab} F^{cd} \partial_{d}F_{hi} \partial_{e}F_{b}{}^{f} 
\partial^{e}F_{ac} \partial^{i}F_{f}{}^{h} - 2 F_{a}{}^{c} F^{ab} 
\partial_{c}F_{hi} \partial_{e}F_{d}{}^{f} \partial^{e}F_{b}{}^{d} 
\partial^{i}F_{f}{}^{h}\nonumber\\
&& -  \frac{1}{4} F_{ab} F^{ab} 
\partial_{e}F_{cd} \partial^{e}F^{cd} \partial_{h}F_{fi} 
\partial^{i}F^{fh} -  \frac{1}{2} F^{ab} F^{cd} 
\partial_{d}F_{ce} \partial^{e}F_{ab} \partial_{i}F_{fh} 
\partial^{i}F^{fh}\nonumber\\
&& -  \frac{1}{2} F^{ab} F^{cd} 
\partial_{e}F_{cd} \partial^{e}F_{ab} \partial_{i}F_{fh} 
\partial^{i}F^{fh} + F^{ab} F^{cd} \partial_{d}F_{be} 
\partial^{e}F_{ac} \partial_{i}F_{fh} \partial^{i}F^{fh} \nonumber\\
&&+ F^{ab} 
F^{cd} \partial_{e}F_{bd} \partial^{e}F_{ac} \partial_{i}F_{fh} 
\partial^{i}F^{fh} -  F_{a}{}^{c} F^{ab} \partial_{c}F_{de} 
\partial^{e}F_{b}{}^{d} \partial_{i}F_{fh} \partial^{i}F^{fh} \nonumber\\
&&+ 
F_{a}{}^{c} F^{ab} \partial_{d}F_{ce} \partial^{e}F_{b}{}^{d} 
\partial_{i}F_{fh} \partial^{i}F^{fh} + 2 F_{a}{}^{c} F^{ab} 
\partial_{e}F_{cd} \partial^{e}F_{b}{}^{d} \partial_{i}F_{fh} 
\partial^{i}F^{fh}\nonumber\\
&& -  \frac{1}{4} F_{ab} F^{ab} 
\partial_{e}F_{cd} \partial^{e}F^{cd} \partial_{i}F_{fh} \partial^{i}F^{fh}.\labell{LF6}
\eeqa
   Because of the presence of the nonlinear $SL(2,R)$ invariant structures in the final S-dual form of the action \reef{LS}, one can derive new couplings at the order of N gauge fields  with four derivatives by expanding the invariant structures. We see that the consistency of the action \reef{LF4} with the S-duality predicts the couplings \reef{LS}
and the couplings at the level of more gauge fields in the Einstein frame. It would be interesting to confirm these couplings by direct calculations\footnote{The S-matrix elements may be constrained by both manifest Lorentz and duality invariance. Considering the duality invariance, as we did in this paper, may be interpreted as an intermediate step towards an analysis of the scattering amplitude \cite{Carrascoa}.}.\\

  {\bf Acknowledgments}: 

 The authors would like to kindly thank M.R.Garousi and A.A.Tseytlin for useful comments and discussions on related topics.\\

{\LARGE \bf Apendix}

In this appendix we are going to find all identities that should be taken in to account when one wants to write the action \reef{LF4} in terms of $SL(2,R)$ invariant structures $WW$. In fact, the terms containing four dual gauge fields and four derivatives should be written in terms of four gauge fields and four derivatives. 

Let us begin with the simple case, where we have four dual gauge fields with zero derivative. There are two different contracted forms for these fields as $ \tilde{F}_{a}{}^{c} \tilde{F}^{ab} \tilde{F}_{b}{}^{d} \tilde{F}_{cd}$ and $\tilde{F}_{ab} \tilde{F}^{ab} \tilde{F}_{cd} \tilde{F}^{cd}$.
Using the identity \reef{ee} one can write any of these terms in terms of four gauge fields contraction $(\tilde{F_1} \tilde{F_2} \tilde{F_3} \tilde{F_4} \mathop{\longrightarrow} \epsilon_{1} \epsilon_{2} \epsilon_{3} \epsilon_{4}  F_{1} F_{2}  F_{3}  F_{4})$. The result in this case is not dependent on the Levi-Civita paring choices $\cE,\cE'$ and $\cE''$.
\beqa
 {\tilde{F}}_{a}{}^{c} {\tilde{F}}^{ab} {\tilde{F}}_{b}{}^{d} {\tilde{F}}_{cd}\mathop{\longrightarrow} {F}_{a}{}^{c} {F}^{ab} {F}_{b}{}^{d} {F}_{cd}\,\,\,\,\,\,\,\,\,\,\,\,\,,\,\,\,\,\,\,\,\,\,{\tilde{F}}_{ab} {\tilde{F}}^{ab} {\tilde{F}}_{cd}{\tilde{F}}^{cd}\mathop{\longrightarrow} F_{ab} F^{ab} F_{cd} F^{cd}.\nonumber
 \eeqa

In the next case where we have four dual gauge fields and two derivatives, the result is dependent on the Levi-Civita paring choices. Consider all contraction of this case with arbitrary coefficients $y_i$ as following:
\beqa
&&y_1 \tilde{F}^{ab} \tilde{F}^{cd} \partial_{b}\tilde{F}_{de} \partial_{c}\tilde{F}_{a}{}^{e} + y_2 \tilde{F}_{a}{}^{c} \tilde{F}^{ab} \partial_{b}\tilde{F}^{de}\partial_{c}\tilde{F}_{de} + y_3 \tilde{F}^{ab} \tilde{F}^{cd} \partial_{c}\tilde{F}_{a}{}^{e} \partial_{d}\tilde{F}_{be} \nonumber\\
&&+y_4 \tilde{F}^{ab} \tilde{F}^{cd}\partial_{b}\tilde{F}_{a}{}^{e} \partial_{d}\tilde{F}_{ce} + y_5 \tilde{F}_{a}{}^{c} \tilde{F}^{ab} \partial_{d}\tilde{F}_{b}{}^{d} \partial_{e}\tilde{F}_{c}{}^{e} + y_6 \tilde{F}^{ab} \tilde{F}^{cd} \partial_{b}\tilde{F}_{ac} \partial_{e}\tilde{F}_{d}{}^{e} \nonumber\\
&&+ y_7 \tilde{F}^{ab} \tilde{F}^{cd} \partial_{c}\tilde{F}_{ab} \partial_{e}\tilde{F}_{d}{}^{e}+ y_8 \tilde{F}_{a}{}^{c} \tilde{F}^{ab} \partial_{c}\tilde{F}_{b}{}^{d} \partial_{e}\tilde{F}_{d}{}^{e} + y_9 \tilde{F}_{ab} \tilde{F}^{ab} \partial_{c}\tilde{F}^{cd} \partial_{e}\tilde{F}_{d}{}^{e}\nonumber\\
&& + y_{10} \tilde{F}^{ab} \tilde{F}^{cd}\partial_{d}\tilde{F}_{ce} \partial^{e}\tilde{F}_{ab} + y_{11} \tilde{F}^{ab} \tilde{F}^{cd} \partial_{e}\tilde{F}_{cd} \partial^{e}\tilde{F}_{ab} + y_{12} \tilde{F}^{ab} \tilde{F}^{cd} \partial_{d}\tilde{F}_{be} \partial^{e}\tilde{F}_{ac}\nonumber\\
&& +y_{13} \tilde{F}^{ab} \tilde{F}^{cd} \partial_{e}\tilde{F}_{bd} \partial^{e}\tilde{F}_{ac} + y_{14} \tilde{F}_{a}{}^{c} \tilde{F}^{ab} \partial_{c}\tilde{F}_{de}\partial^{e}\tilde{F}_{b}{}^{d} + y_{15} \tilde{F}_{a}{}^{c} \tilde{F}^{ab} \partial_{d}\tilde{F}_{ce}\partial^{e}\tilde{F}_{b}{}^{d} \nonumber\\
&&+ y_{16} \tilde{F}_{a}{}^{c} \tilde{F}^{ab} \partial_{e}\tilde{F}_{cd}\partial^{e}\tilde{F}_{b}{}^{d} + y_{17} \tilde{F}_{ab} 
\tilde{F}^{ab} \partial_{d}\tilde{F}_{ce} \partial^{e}\tilde{F}^{cd} +y_{18} \tilde{F}_{ab} \tilde{F}^{ab} \partial_{e}\tilde{F}_{cd} \partial^{e}\tilde{F}^{cd}.\labell{yi}
\eeqa

By selecting the Levi-Civita paring choices $\cE,\cE'$ and $\cE''$, one can find the first term of above dual gauge content in terms of the following different gauge fields combination, respectively:
\beqa
&&- \frac{1}{4} F^{bc} F^{de} \partial_{a}F_{de}\partial^{a}F_{bc} + F^{bc} F^{de} \partial^{a}F_{bc}\partial_{e}F_{da} -  F_{b}{}^{d} F^{bc} \partial^{a}F_{c}{}^{e} 
\partial_{e}F_{da}= Y_1^{},\nonumber\\
 &&  -2 F^{ab} F^{cd} \partial_{b}F_{ac} \partial_{e}F_{d}{}^{e} + F^{ab} F^{cd} \partial_{c}F_{ab} \partial_{e}F_{d}{}^{e} + \frac{1}{4} F^{ab} F^{cd} \partial_{e}F_{cd} \partial^{e}F_{ab}\nonumber\\
&& +2 F^{ab} F^{cd} \partial_{d}F_{be} \partial^{e}F_{ac} -  F^{ab}F^{cd} \partial_{e}F_{bd} \partial^{e}F_{ac} -  F_{a}{}^{c} F^{ab}\partial_{d}F_{ce} \partial^{e}F_{b}{}^{d}=Y_1^{'},\nonumber\\
&&- \frac{1}{2} F_{a}{}^{c} F^{ab} \partial_{b}F^{ed}\partial_{c}F_{ed} - 2 F_{a}{}^{c} F^{ab} \partial_{c}F_{ed} \partial^{d}F_{b}{}^{e} -  F_{a}{}^{c} F^{ab} \partial_{d}F_{ce}\partial^{d}F_{b}{}^{e}\nonumber\\
&& + \frac{1}{4} F_{ab} F^{ab}\partial_{d}F_{ce} \partial^{d}F^{ce} + F^{ab} F^{ce} \partial_{b}F_{a}{}^{d} \partial_{e}F_{cd} -  \frac{1}{2} F_{ab} 
F^{ab} \partial^{d}F^{ce} \partial_{e}F_{cd}= Y_1^{''}.
\eeqa
As $Y_1,Y'_1$ and $Y^{''}_1$ are the expressions for a term containing dual gauge fields in terms of the gauge fields, they should be equal to each other. From $Y_1=Y'_1=Y^{''}_1$, one can find the following  independent identities
\beqa
0&=&4 F^{ab} F^{cd} \partial_{b}F_{ac} \partial_{e}F_{d}{}^{e} -2  F^{ab} F^{cd} \partial_{c}F_{ab} \partial_{e}F_{d}{}^{e} +2 F^{ab}F^{cd} \partial_{d}F_{ce} \partial^{e}F_{ab}\nonumber\\
&& -  F^{ab} F^{cd} \partial_{e}F_{cd} \partial^{e}F_{ab} -4 F^{ab}F^{cd} \partial_{d}F_{be} \partial^{e}F_{ac} +2 F^{ab} F^{cd}\partial_{e}F_{bd} \partial^{e}F_{ac}\,\,,\nonumber\\
0&=&2 F_{a}{}^{c} F^{ab} \partial_{b}F^{de}\partial_{c}F_{de} -4  F^{ab} F^{cd} \partial_{b}F_{a}{}^{e} \partial_{d}F_{ce} +4 F^{ab} F^{cd} \partial_{d}F_{ce}\partial^{e}F_{ab}\nonumber\\
&& -  F^{ab} F^{cd} \partial_{e}F_{cd} \partial^{e}F_{ab} + 8 F_{a}{}^{c} F^{ab} \partial_{c}F_{de} \partial^{e}F_{b}{}^{d} -4  F_{a}{}^{c} F^{ab} 
\partial_{d}F_{ce} \partial^{e}F_{b}{}^{d}\nonumber\\
&& +4 F_{a}{}^{c} F^{ab} \partial_{e}F_{cd} \partial^{e}F_{b}{}^{d} +2 F_{ab} F^{ab} \partial_{d}F_{ce} \partial^{e}F^{cd} -F_{ab} F^{ab} \partial_{e}F_{cd} \partial^{e}F^{cd}.\labell{idp1}
\eeqa
To write the second term in \reef{yi} in terms of the gauge fields, one finds three expressions $Y_2,Y'_2$ and $Y^{''}_2$ according to the Levi-Civita paring choices $\cE,\cE'$ and $\cE''$. Considering the relation $Y_2=Y'_2=Y^{''}_2$ and the identities \reef{idp1}, the another independent identity appears as following: 
\beqa
0&=&4 F^{ab} F^{cd} \partial_{b}F_{de} \partial_{c}F_{a}{}^{e} +2  F_{a}{}^{c} F^{ab} \partial_{b}F^{de} \partial_{c}F_{de} -4 F_{a}{}^{c} F^{ab} \partial_{d}F_{b}{}^{d} \partial_{e}F_{c}{}^{e} \nonumber\\
&&-8 F^{ab} F^{cd} \partial_{b}F_{ac} \partial_{e}F_{d}{}^{e} +8 F_{a}{}^{c} F^{ab} \partial_{c}F_{b}{}^{d} \partial_{e}F_{d}{}^{e} -2 F_{ab} F^{ab} \partial_{c}F^{cd} \partial_{e}F_{d}{}^{e}\nonumber\\
&& -4F^{ab} F^{cd} \partial_{d}F_{ce} \partial^{e}F_{ab} + F^{ab} F^{cd} \partial_{e}F_{cd} \partial^{e}F_{ab} +8 F^{ab} F^{cd} \partial_{d}F_{be} \partial^{e}F_{ac} \nonumber\\
&&-4F^{ab} F^{cd} \partial_{e}F_{bd} \partial^{e}F_{ac} +4 F_{a}{}^{c} F^{ab} \partial_{e}F_{cd} \partial^{e}F_{b}{}^{d} - F_{ab} 
F^{ab} \partial_{e}F_{cd} \partial^{e}F^{cd}.
\eeqa
Doing the same procedure for the third term in \reef{yi} led to the following identity that is independent of previous identities.
\beqa
0&=& 2 F^{ac} F^{de} \partial_{d}F_{a}{}^{b} \partial_{e}F_{cb} -2 F_{a}{}^{d} F^{ac} \partial_{e}F_{c}{}^{e} \partial_{b}F_{d}{}^{b} -2 F^{ac} F^{de} \partial_{c}F_{ad} \partial_{b}F_{e}{}^{b} \nonumber\\
&& - F^{ac} F^{de} \partial_{d}F_{ac} \partial_{b}F_{e}{}^{b} +4 F_{a}{}^{d} F^{ac} \partial_{d}F_{c}{}^{e} \partial_{b}F_{e}{}^{b} -F_{ac} F^{ac} \partial_{d}F^{de} \partial_{b}F_{e}{}^{b}\nonumber\\
&&  - F^{ac} F^{de} \partial_{e}F_{db} \partial^{b}F_{ac}-2 F^{ac} F^{de} \partial_{e}F_{cb} \partial^{b}F_{ad} -4 F_{a}{}^{d} F^{ac} \partial_{d}F_{eb} \partial^{b}F_{c}{}^{e}\nonumber\\
&&  +2  F_{a}{}^{d}F^{ac} \partial_{e}F_{db} \partial^{b}F_{c}{}^{e} - F_{ac} F^{ac} \partial_{e}F_{db} \partial^{b}F^{de}.
\eeqa
After finding the other terms in \reef{yi} in terms of the gauge fields ($Y_i,Y^{'}_i$ and $Y^{''}_i$ where $i=3,\cdots,18$), one finds that the relations $Y_i=Y^{'}_i=Y^{''}_i$ do not produce any new independent indentity. 

Let us now consider the third case where we have four gauge fields and four derivatives. One can find all contracted form of four dual gauge fields with four derivatives by using "xAct" as following:
\beqa
&&z_1 \partial_{c}\tilde{F}^{ef} \partial^{c}\tilde{F}^{ab} \partial_{d}\tilde{F}_{ef} \partial^{d}\tilde{F}_{ab} + z_2 \partial_{a}\tilde{F}^{ab} \partial_{c}\tilde{F}^{ef} \partial_{d}\tilde{F}_{ef} \partial^{d}\tilde{F}_{b}{}^{c} + z_3 \partial_{b}\tilde{F}_{ef} \partial^{c}\tilde{F}^{ab}\partial_{d}\tilde{F}_{c}{}^{f} \partial^{e}\tilde{F}_{a}{}^{d}\nonumber\\
&&+ z_4 \partial_{b}\tilde{F}_{c}{}^{f} \partial^{c}\tilde{F}^{ab} \partial_{d}\tilde{F}_{ef} \partial^{e}\tilde{F}_{a}{}^{d} + z_5\partial_{c}\tilde{F}_{df} \partial^{c}\tilde{F}^{ab} \partial_{e}\tilde{F}_{b}{}^{f} \partial^{e}\tilde{F}_{a}{}^{d} + z_6 \partial_{a}\tilde{F}^{ab} \partial_{c}\tilde{F}_{b}{}^{c} \partial_{d}\tilde{F}^{de} \partial_{f}\tilde{F}_{e}{}^{f} \nonumber\\
&&+ z_7 \partial_{a}\tilde{F}^{ab} \partial_{c}\tilde{F}_{d}{}^{e} \partial^{d}\tilde{F}_{b}{}^{c} \partial_{f}\tilde{F}_{e}{}^{f} + z_8
\partial_{a}\tilde{F}^{ab} \partial_{d}\tilde{F}_{c}{}^{e} \partial^{d}\tilde{F}_{b}{}^{c} \partial_{f}\tilde{F}_{e}{}^{f} + z_9 \partial_{a}\tilde{F}^{ab} 
\partial_{b}\tilde{F}^{cd} \partial^{e}\tilde{F}_{cd} \partial_{f}\tilde{F}_{e}{}^{f}  \nonumber\\
&&+z_{10} \partial_{a}\tilde{F}^{ab} \partial^{d}\tilde{F}_{b}{}^{c} \partial^{e}\tilde{F}_{cd} \partial_{f}\tilde{F}_{e}{}^{f} + z_{11}
\partial^{c}\tilde{F}^{ab} \partial^{e}\tilde{F}_{a}{}^{d} \partial_{f}\tilde{F}_{ce} \partial^{f}\tilde{F}_{bd} + z_{12}\partial_{c}\tilde{F}_{df} 
\partial^{c}\tilde{F}^{ab} \partial^{e}\tilde{F}_{a}{}^{d} \partial^{f}\tilde{F}_{be}\nonumber\\
&& + z_{13} \partial^{c}\tilde{F}^{ab} \partial_{d}\tilde{F}_{cf} \partial^{e}\tilde{F}_{a}{}^{d} \partial^{f}\tilde{F}_{be} + z_{14}
\partial^{c}\tilde{F}^{ab} \partial^{e}\tilde{F}_{a}{}^{d} \partial_{f}\tilde{F}_{cd} \partial^{f}\tilde{F}_{be} + z_{15} \partial^{c}\tilde{F}^{ab} 
\partial_{d}\tilde{F}_{ef} \partial^{d}\tilde{F}_{ac} \partial^{f}\tilde{F}_{b}{}^{e}\nonumber\\
&& + z_{16} \partial_{c}\tilde{F}_{a}{}^{d} \partial^{c}\tilde{F}^{ab} \partial_{e}\tilde{F}_{df} \partial^{f}\tilde{F}_{b}{}^{e} + z_{17}
\partial^{c}\tilde{F}^{ab} \partial^{d}\tilde{F}_{ac} \partial_{e}\tilde{F}_{df} \partial^{f}\tilde{F}_{b}{}^{e} + z_{18}\partial_{c}\tilde{F}_{a}{}^{d} 
\partial^{c}\tilde{F}^{ab} \partial_{f}\tilde{F}_{de} \partial^{f}\tilde{F}_{b}{}^{e}\nonumber\\
&& + z_{19} \partial^{c}\tilde{F}^{ab} \partial^{d}\tilde{F}_{ac} \partial_{f}\tilde{F}_{de} \partial^{f}\tilde{F}_{b}{}^{e} + z_{20}
\partial_{a}\tilde{F}^{ab} \partial_{b}\tilde{F}^{cd} \partial_{d}\tilde{F}_{ef} \partial^{f}\tilde{F}_{c}{}^{e} + z_{21} \partial^{c}\tilde{F}^{ab} 
\partial_{d}\tilde{F}_{ef} \partial^{d}\tilde{F}_{ab} \partial^{f}\tilde{F}_{c}{}^{e}\nonumber\\
&& + z_{22} \partial_{a}\tilde{F}^{ab} \partial_{d}\tilde{F}_{ef} \partial^{d}\tilde{F}_{b}{}^{c} \partial^{f}\tilde{F}_{c}{}^{e} + z_{23}
\partial^{c}\tilde{F}^{ab} \partial^{d}\tilde{F}_{ab} \partial_{e}\tilde{F}_{df} \partial^{f}\tilde{F}_{c}{}^{e} + z_{24}\partial_{a}\tilde{F}^{ab} 
\partial^{d}\tilde{F}_{b}{}^{c} \partial_{e}\tilde{F}_{df} \partial^{f}\tilde{F}_{c}{}^{e} \nonumber\\
&&+ z_{25} \partial^{c}\tilde{F}^{ab} \partial^{d}\tilde{F}_{ab} \partial_{f}\tilde{F}_{de} \partial^{f}\tilde{F}_{c}{}^{e} + z_{26}
\partial_{a}\tilde{F}^{ab} \partial^{d}\tilde{F}_{b}{}^{c} \partial_{f}\tilde{F}_{de} \partial^{f}\tilde{F}_{c}{}^{e} + z_{27} \partial_{b}\tilde{F}_{ef} 
\partial^{c}\tilde{F}^{ab} \partial^{d}\tilde{F}_{ac} \partial^{f}\tilde{F}_{d}{}^{e} \nonumber\\
&& + z_{28} \partial_{a}\tilde{F}^{ab} \partial_{c}\tilde{F}_{ef} \partial^{d}\tilde{F}_{b}{}^{c} \partial^{f}\tilde{F}_{d}{}^{e}+ z_{29}
\partial_{a}\tilde{F}^{ab} \partial_{c}\tilde{F}_{b}{}^{c} \partial_{e}\tilde{F}_{df} \partial^{f}\tilde{F}^{de} + z_{30} \partial_{b}\tilde{F}_{ac} 
\partial^{c}\tilde{F}^{ab} \partial_{e}\tilde{F}_{df} \partial^{f}\tilde{F}^{de}\nonumber\\
&&  + z_{31} \partial_{c}\tilde{F}_{ab} \partial^{c}\tilde{F}^{ab} \partial_{e}\tilde{F}_{df} \partial^{f}\tilde{F}^{de} + z_{32}
\partial_{a}\tilde{F}^{ab} \partial_{c}\tilde{F}_{b}{}^{c} \partial_{f}\tilde{F}_{de} \partial^{f}\tilde{F}^{de} + z_{33} \partial_{c}\tilde{F}_{ab} 
\partial^{c}\tilde{F}^{ab} \partial_{f}\tilde{F}_{de} \partial^{f}\tilde{F}^{de},\nonumber
\eeqa
where $z_i$ are some unknown coefficients.

As the previous cases, we do the same procedure to find the above dual gauge fields in terms of relevant gauge fields. 
Different choices of Levi-Civita paring led to different results that shoud be equal to each other  $Z_i=Z^{'}_i=Z^{''}_i$ and $i=1,\cdots,33$. These equalities led to ten independent identities which we present three of them in the following. To investigate the S-duality behavior of amplitude and actions at the presence of a $D$-brane at the level of four gauge fields and four(/N) derivatives, one has to use these identities as we have done to write the action \reef{LF4} in terms of the S-dual invariant structures \reef{LS}. 
\beqa
0&=& \partial_{c}F^{ef} \partial^{c}F^{ab} \partial_{d}F_{ef} \partial^{d}F_{ab} -2 \partial_{c}F_{df} \partial^{c}F^{ab} \partial_{e}F_{b}{}^{f}\partial^{e}F_{a}{}^{d} -4 \partial^{c}F^{ab} \partial^{e}F_{a}{}^{d} \partial_{f}F_{ce} \partial^{f}F_{bd} \nonumber\\
&& +8  \partial_{c}F_{df} \partial^{c}F^{ab} \partial^{e}F_{a}{}^{d} \partial^{f}F_{be} +4 \partial^{c}F^{ab} \partial^{e}F_{a}{}^{d} \partial_{f}F_{cd} \partial^{f}F_{be} -4 \partial^{c}F^{ab} \partial_{d}F_{ef} \partial^{d}F_{ac} \partial^{f}F_{b}{}^{e} \nonumber\\
&& +4 \partial^{c}F^{ab} \partial_{d}F_{ef} \partial^{d}F_{ab} \partial^{f}F_{c}{}^{e}\,\,\,,\nonumber\\
0&=& \partial_{c}F^{ef} \partial^{c}F^{ab} \partial_{d}F_{ef} \partial^{d}F_{ab} -4 \partial^{c}F^{ab} \partial_{d}F_{ef} \partial^{d}F_{ac} \partial^{f}F_{b}{}^{e} +4  \partial_{c}F_{a}{}^{d} \partial^{c}F^{ab} \partial_{e}F_{df} \partial^{f}F_{b}{}^{e}  \nonumber\\
&&-4 \partial_{c}F_{a}{}^{d} \partial^{c}F^{ab} \partial_{f}F_{de} \partial^{f}F_{b}{}^{e} +8 \partial^{c}F^{ab} \partial^{d}F_{ac} \partial_{f}F_{de} 
\partial^{f}F_{b}{}^{e} +4  \partial^{c}F^{ab} \partial_{d}F_{ef} \partial^{d}F_{ab} \partial^{f}F_{c}{}^{e}   \nonumber\\
&&-2 \partial^{c}F^{ab} \partial^{d}F_{ab} \partial_{f}F_{de} \partial^{f}F_{c}{}^{e}-2 \partial_{c}F_{ab}\partial^{c}F^{ab} \partial_{e}F_{df} \partial^{f}F^{de} +  \partial_{c}F_{ab} \partial^{c}F^{ab} \partial_{f}F_{de} \partial^{f}F^{de}\,\,\,,\nonumber\\
0&=& \partial_{c}F^{ef} \partial^{c}F^{ab} \partial_{d}F_{ef} \partial^{d}F_{ab}+4 \partial_{a}F^{ab} \partial_{c}F^{ef} \partial_{d}F_{ef} 
\partial^{d}F_{b}{}^{c} -4 \partial_{c}F_{df} \partial^{c}F^{ab} \partial_{e}F_{b}{}^{f} \partial^{e}F_{a}{}^{d}  \nonumber\\
&&+8 \partial_{c}F_{df} \partial^{c}F^{ab} \partial^{e}F_{a}{}^{d} \partial^{f}F_{be}+4 \partial^{c}F^{ab} \partial_{d}F_{ef} \partial^{d}F_{ac} \partial^{f}F_{b}{}^{e} -4 \partial_{c}F_{a}{}^{d} \partial^{c}F^{ab} \partial_{e}F_{df} \partial^{f}F_{b}{}^{e} \nonumber\\
&& +4 \partial_{c}F_{a}{}^{d} \partial^{c}F^{ab} \partial_{f}F_{de} \partial^{f}F_{b}{}^{e}-8 \partial^{c}F^{ab} \partial^{d}F_{ac} \partial_{f}F_{de} \partial^{f}F_{b}{}^{e}  +8 \partial_{a}F^{ab} \partial_{d}F_{ef} \partial^{d}F_{b}{}^{c} \partial^{f}F_{c}{}^{e}\  \nonumber\\
&&+2 \partial^{c}F^{ab} \partial^{d}F_{ab} \partial_{f}F_{de} \partial^{f}F_{c}{}^{e}+2 \partial_{c}F_{ab} \partial^{c}F^{ab} \partial_{e}F_{df} \partial^{f}F^{de} - \partial_{c}F_{ab} \partial^{c}F^{ab} \partial_{f}F_{de} \partial^{f}F^{de}\,\,\,.\nonumber
\eeqa
These identities and the other ones that appear in this section could be checked explicitly by replacing $F_{ab}=\partial_a A_{b}-\partial_b A_{a}$ (or by
substituting the component of $F_{ab}$).

\end{document}